\documentclass[useAMS,usenatbib]{mn2e}

\usepackage{color}
\usepackage{graphics}
\usepackage{psfig,subfigure}
 
\title[Rotationally Modulated X-ray Emission from T Tauri Stars]
      {Rotationally Modulated X-ray Emission from T Tauri Stars}
\author[S. G. Gregory et al.]
{S. G. Gregory$^{1}$\thanks{E-mail: sg64@st-andrews.ac.uk},
 M. Jardine$^{1}$, A. Collier Cameron$^{1}$ and J.- F. Donati$^{2}$ \\
$^{1}$School of Physics and Astronomy, University of St Andrews, North 
Haugh, St Andrews, Fife, KY16 9SS, U. K.\\
$^{2}$Laboratoire d'Astrophysique, Observatoire Midi-Pyr\'en\'ees, 
      14 Av. E. Belin, F-31400 Toulouse, France}

\begin{document}

\date{}

\pagerange{\pageref{firstpage}--\pageref{lastpage}} \pubyear{2006}

\maketitle

\label{firstpage}

\begin{abstract}
We have modelled the rotational modulation of X-ray emission from T Tauri stars assuming that
they have isothermal, magnetically confined coronae.  By extrapolating surface magnetograms we 
find that T Tauri coronae are compact and clumpy, such that rotational modulation arises from 
X-ray emitting regions being eclipsed as the star rotates.  Emitting regions are close to the 
stellar surface and inhomogeneously distributed about the star.  However some 
regions of the stellar surface, which contain wind bearing open field lines, 
are dark in X-rays.  
From simulated X-ray light curves, obtained using stellar parameters from the 
Chandra Orion Ultradeep Project, we calculate X-ray periods and make comparisons with optically
determined rotation periods.  We find that X-ray periods are typically equal to, or are half of,
the optical periods.  Further, we find that X-ray periods are dependent upon the stellar inclination, 
but that the ratio of X-ray to optical period is independent of stellar mass and radius.  
\end{abstract}



\section{Introduction}
One of the recent results from the Chandra Orion Ultradeep Project (COUP) was the detection
of significant rotational modulation of X-ray emission from low mass pre-main 
sequence stars.  The detection of such modulation suggests that the 
coronae of T Tauri stars are compact and clumpy, with emitting regions
that are inhomogeneously distributed across the stellar surface, and confined within 
magnetic structures that do not extend out to much beyond a stellar radius \citep{fla05}.
There is also evidence for much larger magnetic loops, possibly due to the interaction 
with a surrounding circumstellar disc \citep{fav05,gia06}.  A model already exists for T Tauri coronae, 
where complex magnetic field structures contain X-ray emitting plasma close to the stellar 
surface, whilst larger magnetic loops and open field lines are able to carry accretion flows
(\citealt{jar06}; \citealt{gre06}).

In this paper we use surface magnetograms derived from Zeeman-Doppler imaging to extrapolate the 
coronae of T Tauri stars 
using stellar parameters taken from the COUP dataset \citep{get05}.  By considering isothermal 
coronae in hydrostatic equilibrium we obtain simulated X-ray light curves, for a range of stellar 
inclinations, and then X-ray periods using the Lomb Normalised Periodogram (LNP) method.
We compare our results with those of \citet{fla05}, who demonstrate that those COUP stars which show clear
evidence for rotationally modulated X-ray emission appear to have X-ray periods which are
either equal to the optically determined rotation period ($P_X=P_{opt}$) or are half of it 
($P_X=0.5P_{opt}$).        


\begin{figure*}
        \def\subfigtopskip{4pt}
        \def\subfigbottomskip{4pt}
        \def\subfigcapskip{2pt}
        \centering
        \begin{tabular}{cc}
        \subfigure[]{
                        \label{lqhya_extrap}
                        \psfig{figure=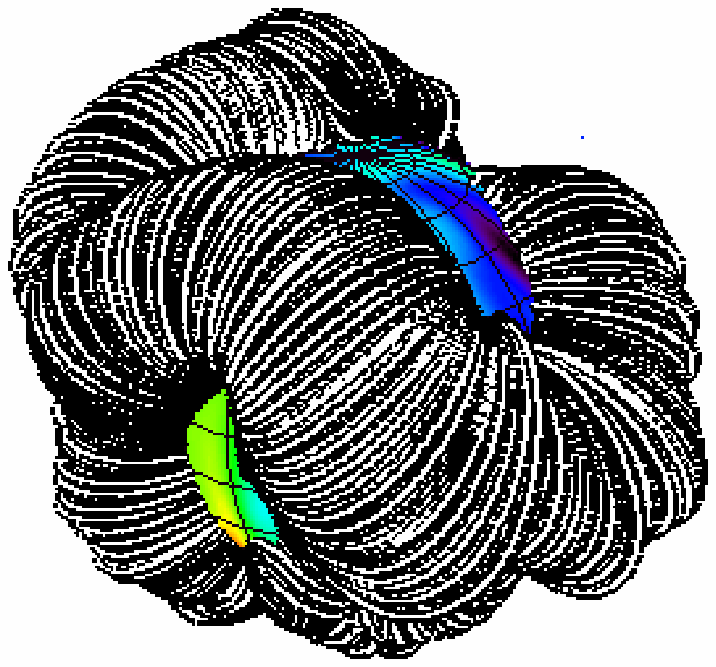,width=80mm}
                        } &
                \subfigure[]{
                        \label{abdor_extrap}
                        \psfig{figure=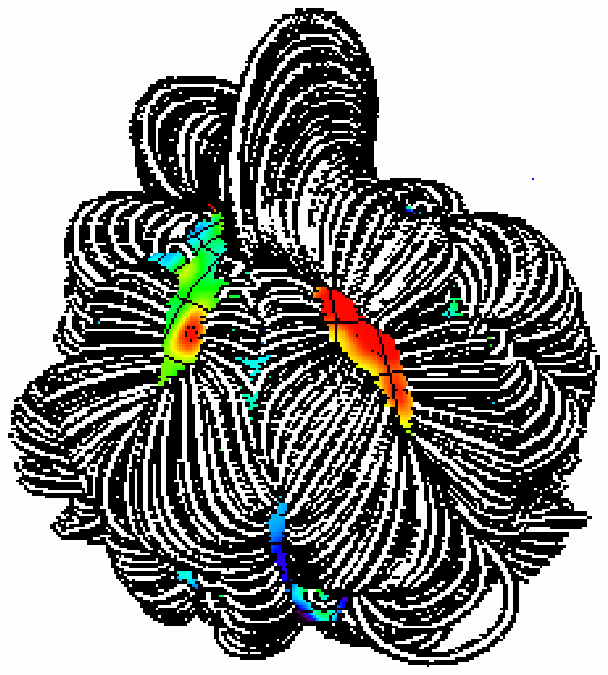,width=80mm}
                        }  \\
        \end{tabular}
        \caption[]{Closed coronal structures showing field lines which contain X-ray emitting plasma, 
                   extrapolated from surface magnetograms of (a) LQ Hya and (b) AB Dor, for COUP 
                   source number 6 ($M_{\ast}=0.23M_{\odot}, R_{\ast}=1.6R_{\odot}, P_{rot}=9.81d$).
                   The magnetic structures are compact and inhomogeneously distributed about the stellar
                   surface.}
        \label{extrap}
\end{figure*}

\section{Realistic magnetic fields}
In order to model the coronae of T Tauri stars we need to assume something about the form of the magnetic field.
Observations suggest it is compact and inhomogeneous and may vary not only with time on each star, but also
from one star to the next. To capture this behaviour, we use as examples the field structures of two different
main sequence stars, LQ Hya and AB Dor determined using Zeeman-Doppler imaging (\citealt{don97a}; \citealt{don97b}; 
\citealt{don99a}; \citealt{don99b}; \citealt{don03}).  By using both stars, we can assess the degree to which 
variations in the detailed structure of a star's corona may affect its X-ray emission.  The method for extrapolating 
the field follows that employed by \citet{jar02a} and is further detailed by \citet{jar06} 
and \citet{gre06}; thus we provide only an 
outline here.  Assuming the magnetic field $\bmath{B}$ is potential then 
$\nabla \times \bmath{B} = 0$. This condition is satisfied by writing the field in terms of a
scalar flux function $\Psi$, such that $\bmath{B}=-\nabla \Psi$.  Thus in order to ensure that the 
field is divergence-free ($\nabla.\bmath{B}=0$), $\Psi$ must satisfy Laplace's equation, 
$\nabla^2\Psi=0$; the solution of which is a linear combination of spherical harmonics,
\begin{equation}
\Psi= \sum_{l=1}^{N} \sum_{m=-l}^{l} \left
[a_{lm}r^l+b_{lm}r^{-(l+1)} \right ] P_{lm}(\theta) {\rm e}^{{\rm i}m\phi},
\end{equation}
where $P_{lm}$ denote the associated Legendre functions.  The coefficients $a_{lm}$ and $b_{lm}$ 
are determined from the radial field at the stellar surface obtained from Zeeman-Doppler maps and 
also by assuming that at some height $R_s$ above the surface (known as the source surface) the 
field becomes radial and hence $B_{\theta}(R_s)=0$, emulating the effect of the corona blowing 
open field lines to form a stellar wind \citep{alt69}.  In order to extrapolate the field we used 
a modified version of a code originally developed by \citet*{van98}.

For a given surface magnetogram we calculate the extent of the closed corona for a particular set
of stellar parameters.  This is achieved by calculating the hydrostatic pressure at each point along 
a field line loop.  For an isothermal corona and assuming that the plasma along the field is in 
hydrostatic equilibrium then,
\begin{equation}
  p=p_0\exp{\left(\frac{1}{c_s^2}\oint_{s}g_sds \right)},
  \label{phydro}
\end{equation}
where $c_s$ is the isothermal sound speed and $g_s$ the component of the effective
gravity along the field line such that, $g_s=\bmath{g}.\bmath{B}/|\bmath{B}|$.  The constant $p_0$
is the gas pressure at a field line foot point and $p$ the pressure at some point along
the field line.  The effective gravity in spherical coordinates for a star with
rotation rate $\omega$ is,
\begin{equation}
  \bmath{g}\left( r,\theta,\phi \right) =
         \left ( -\frac{GM_{\ast}}{r^2}+\omega^2r\sin^2{\theta},
         \omega^2r\sin{\theta}\cos{\theta},0 \right).
  \label{gravity}
\end{equation}
We can then calculate how the plasma $\beta$, the ratio of gas to magnetic pressure, 
changes along each field line.  If at any point along a field line $\beta >1$ then we
assume that the field line is blown open.  This effect is incorporated into our model by 
setting the coronal (gas) pressure to zero whenever it exceeds the magnetic 
pressure ($\beta >1$).  We also set the coronal pressure to zero for open field 
lines.  The gas pressure, and therefore the plasma $\beta$, is dependent upon the choice of 
$p_0$ which is a free parameter of our model.  If we assume that the base pressure is 
proportional to the magnetic pressure then $p_0 = KB_0^2$, a technique which has been used 
successfully to calculate mean coronal densities and X-ray emission measures for the Sun and 
other main sequence stars (\citealt{jar02a,jar02b}) as well as T Tauri stars \citep{jar06}.  
By varying the constant $K$ we can raise or lower the overall gas pressure along field line loops.  
If the value of $K$ is large, many field lines would be blown open and the corona would be 
compact, whilst if the value of $K$ is small, then the magnetic field is able to contain more 
of the coronal gas.  The extent of the corona therefore depends both on the value of $K$ 
and also on $B_0$ which is determined directly from surface magnetograms.  For an observed 
surface magnetogram the base magnetic pressure $B_0$ varies across the stellar surface, and as 
such so does the base pressure $p_0$ at field line foot points.  By considering stars from the 
COUP dataset \citet{jar06} obtain the value of $K$ which results in the best fit to 
observed X-ray emission measures, for a given surface magnetogram.  We have adopted 
the same values in this paper.  We note that we make a conservative estimate of the location of
the source surface by calculating the largest radial distance at which a dipole field line would remain 
closed, with the same average field strength.  Fig. \ref{extrap} shows examples 
of the complex magnetic field geometries which we consider.

The coronal field geometries which we consider in this work have been extrapolated from
surface magnetograms of the young main sequence stars AB Dor and LQ Hya.  In future it will
be possible to use real T Tauri magnetograms dervied from Zeeman-Doppler images obtained using
the ESPaDOnS instrument at the Canada-France-Hawaii telescope.  However, in the meantime, 
the example field geometries shown in Fig. \ref{extrap} capture three essential features
of T Tauri coronae.  First, \citet{jar06} have already demonstrated that the field structures which
we consider here yield X-ray emission measures and mean coronal densities which are consistent 
with values obtained during the COUP - the largest available dataset of X-ray properties of young
stellar objects \citep{get05}.  Second, the fact that rotational modulation of X-ray emission was at all
detected during the COUP automatically led to the conclusion that the dominant X-ray emitting regions
must be compact and unevenly distributed across the star \citep{fla05}.  It can be seen from 
Fig. \ref{extrap} that the X-ray emitting closed coronal field lines do not extent out to much beyond the stellar
surface (the coronae are compact) and that the emitting regions are inhomogeneously distributed about the star
(the coronae are clumpy), in immediate agreement with the COUP results.  Further, as we discuss below, 
the field geometries considered in this paper do give rise to significant rotational 
modulation of X-ray emission.  It can also be seen from Fig. \ref{extrap} that some regions of the stellar
surface do not contain regions of closed field, and would therefore be dark in X-rays.  Such regions 
contain wind bearing open field lines.  Third, observations of various classical T Tauri stars by \citet{val04} 
show the line-of-sight (longitudinal) magnetic field components measured using photospheric absorption lines,
are often consistent with a net circular polarisation signal of zero.  However, strong
fields are detected from Zeeman broadening measurements.  Such polarisation measurements trace the surface 
field structure of T Tauri stars and immediately imply that such stars have complex and highly structured 
coronae.  If the surface of the star is covered in many regions of opposite polarity this would give rise
to a net polarisation signature of zero, with contributions to the overall signature from regions
of opposite polarity cancelling out.  Therefore the surface field must be highly complex and multipolar in nature,
with many closed loops confining X-ray emitting plasma close to the surface of the star.  However, it is also worth 
asking whether the coronal structure of classical T Tauri stars are likely to be fundamentally different from those of 
weak line T Tauri stars, a point which we discuss in \S4.

Although we cannot be certain whether or not the magnetic field structures extrapolated from 
surface magnetograms of young main sequence stars in Fig. \ref{extrap} do
represent the magnetically confined coronae of T Tauri stars, they do satisfy the currently available
observational constraints.  They reproduce X-ray emission measures and coronal densities which 
are typical of T Tauri stars, and the field structures are complex, confining plasma within unevenly
distributed magnetic structures close to the stellar surface.  However, as such structures 
only represent a snap-shot in time of the coronal field geometry, we have chosen to consider two
different field topologies.  This allows us to model how much of an effect the field geometry has
on the amplitude of modulation of X-ray emission and on resulting X-ray periods.


\section{X-ray periods}

\subsection{X-ray light curves}

\begin{figure*}
        \def\subfigtopskip{4pt}
        \def\subfigbottomskip{4pt}
        \def\subfigcapskip{2pt}
        \centering
        \begin{tabular}{cc}
            \subfigure[]{
                        \label{lqhya_lightcurve}
                        \psfig{figure=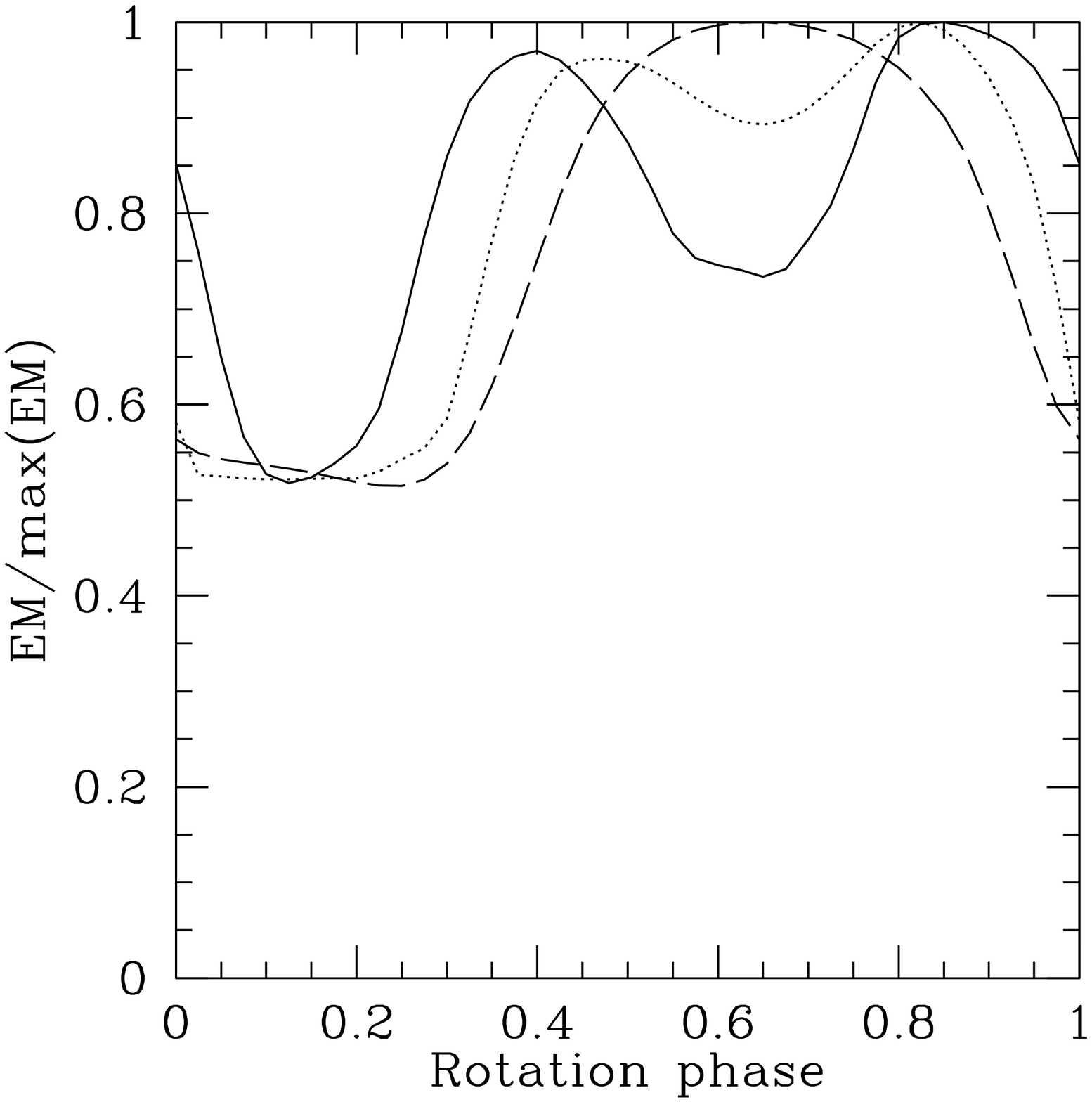,width=80mm}
                        } &
            \subfigure[]{
                        \label{abdor_lightcurve}
                        \psfig{figure=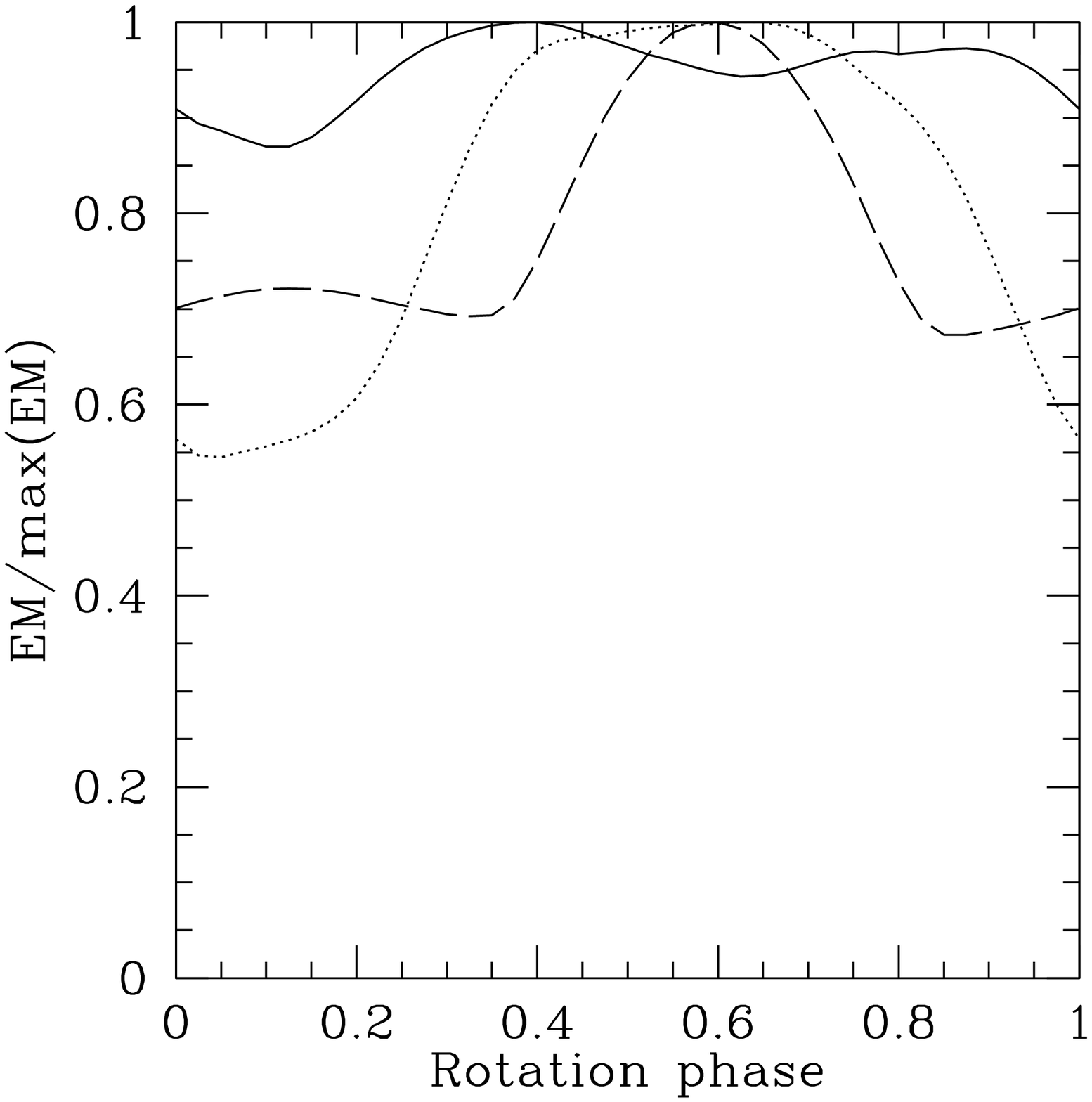,width=80mm}
                        }  \\
        \end{tabular}
        \caption[]{The variation in X-ray emission measure (EM) with rotation phase for (a) the LQ Hya-like and (b) the AB Dor-like 
                  coronal structures as shown in Fig. \ref{extrap}, for inclinations of 30\degr({\it dashed}), 
                  60\degr({\it dotted}) and 90\degr({\it solid}).  There is clear rotational modulation of X-ray 
                  emission.}
        \label{lightcurve}
\end{figure*}

\begin{figure*}
        \def\subfigtopskip{4pt}
        \def\subfigbottomskip{4pt}
        \def\subfigcapskip{2pt}
        \centering
        \begin{tabular}{cccc}
            \subfigure[]{
                        \label{lqhya_frame1}
                        \psfig{figure=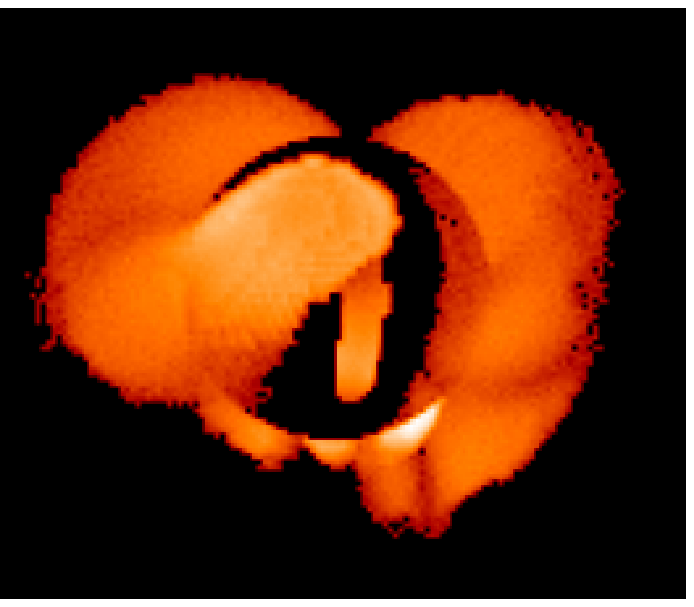,width=35mm}
                        } &
            \subfigure[]{
                        \label{lqhya_frame2}
                        \psfig{figure=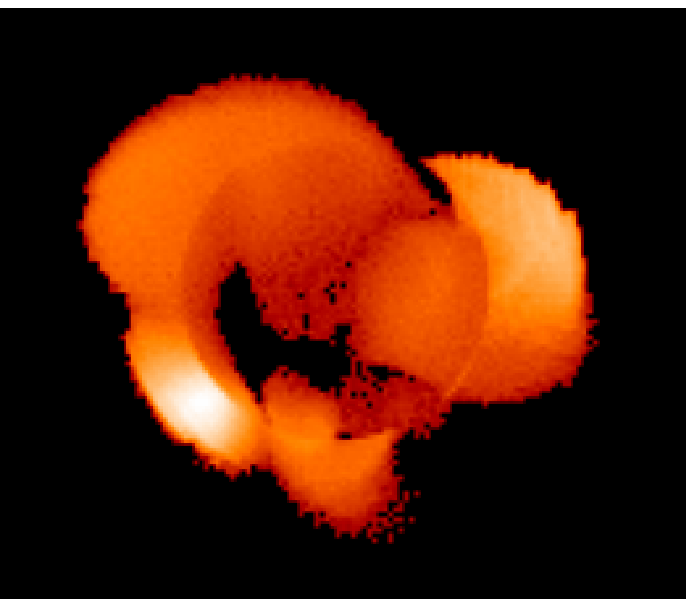,width=35mm}
                        } &
            \subfigure[]{
                        \label{lqhya_frame3}
                        \psfig{figure=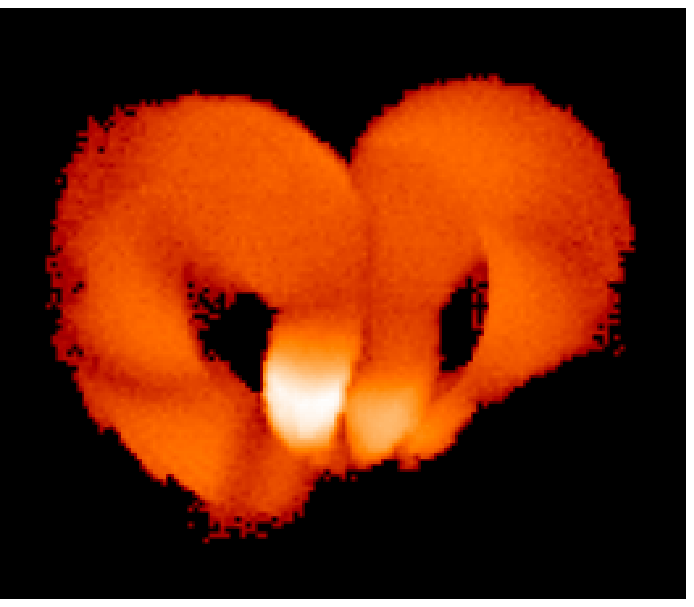,width=35mm}
                        } &
            \subfigure[]{
                        \label{lqhya_frame4}
                        \psfig{figure=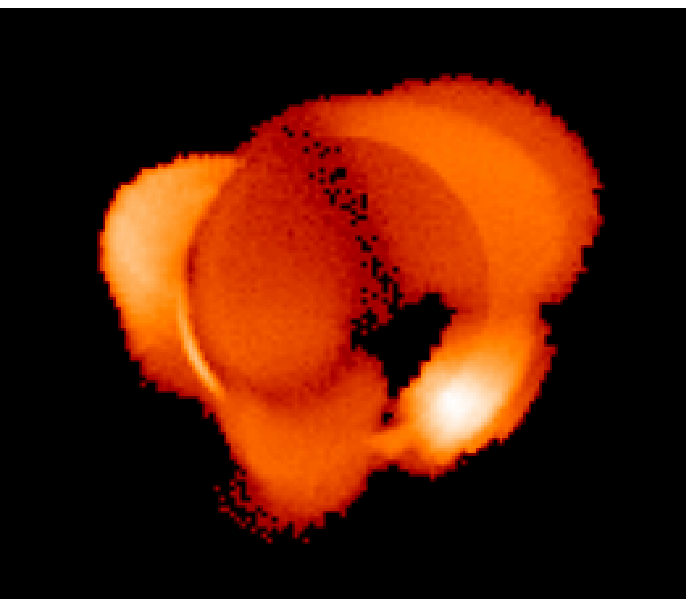,width=35mm}
                        }  \\
        \end{tabular}
        \caption[]{X-ray images obtained using the LQ Hya-like coronal structure from Fig. \ref{lqhya_extrap} 
                  for an inclination 90\degr (shown as the {\it solid line} in Fig. \ref{lqhya_lightcurve}).  
		  Emitting regions are 
                  inhomogeneously distributed across the stellar surface and are confined within magnetic structures close 
                  to the star; some regions are, however, dark in X-rays.  This particular surface map has two dominant 
                  emitting regions in opposite hemispheres, which at this inclination, go into eclipse as the star rotates
                  giving rise to the two minima in Fig. \ref{lqhya_lightcurve} ({\it solid line}).  Image (a) is for a 
                  rotational phase of 0.1, where the brightest of the two dominant emitting regions is in eclipse, (b) 0.4, where
                  both of the dominant regions are visible, (c) 0.65, where the brightest emitting region is visible and the other
                  is in eclipse and (d) 0.85, where once again both of the dominant regions can be seen.}
        \label{lqhya}
\end{figure*}

\begin{figure*}
        \def\subfigtopskip{4pt}
        \def\subfigbottomskip{4pt}
        \def\subfigcapskip{2pt}
        \centering
        \begin{tabular}{cccc}
            \subfigure[]{
                        \label{abdor_frame1}
                        \psfig{figure=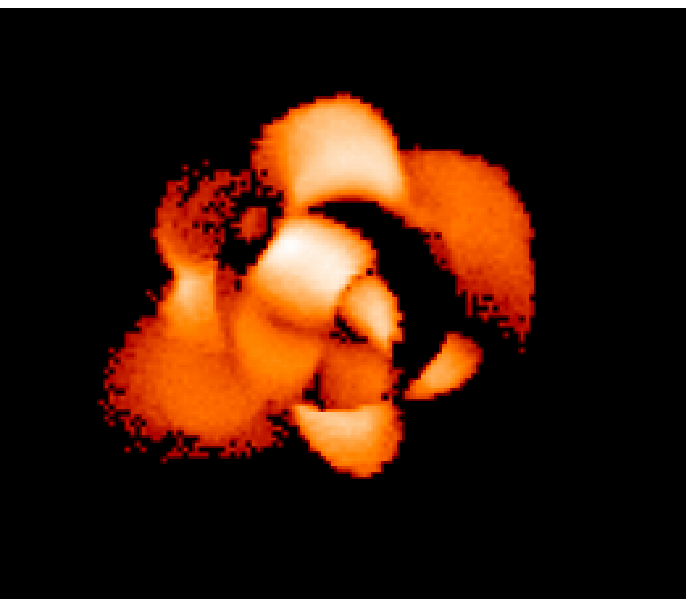,width=35mm}
                        } &
            \subfigure[]{
                        \label{abdor_frame2}
                        \psfig{figure=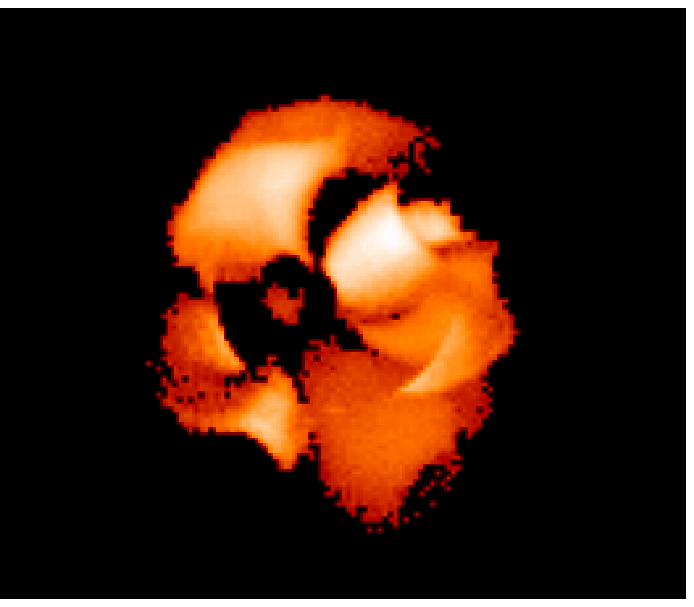,width=35mm}
                        } &
            \subfigure[]{
                        \label{abdor_frame3}
                        \psfig{figure=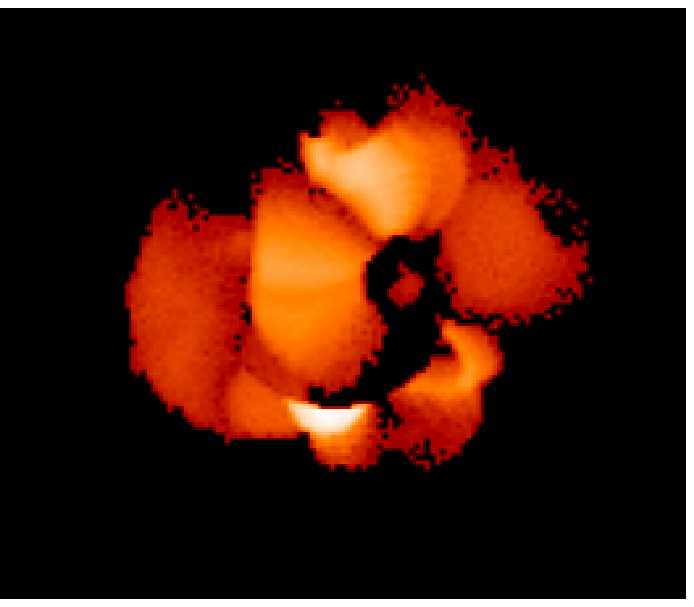,width=35mm}
                        } &
            \subfigure[]{
                        \label{abdor_frame4}
                        \psfig{figure=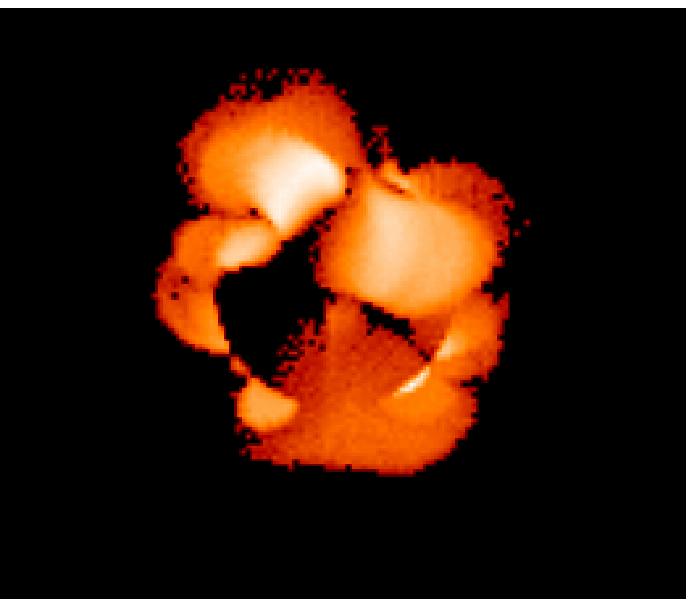,width=35mm}
                        }  \\
        \end{tabular}
        \caption[]{As Fig. \ref{lqhya} but using the AB Dor-like coronal structure from Fig. \ref{abdor_extrap} and 
                   for an inclination of 30\degr (shown as the dashed line in Fig. \ref{abdor_lightcurve}).  
                   Image (a) is for a rotational phase of 0.15, (b) 0.35, (c) 0.6, where the
                   amount of X-ray emission is at a maximum due to the brightest emitting region (towards the bottom 
                   of the image) being visible and (d) 0.85.  The coronal structure is more clumpy than the LQ Hya-like field, 
                   with many bright emitting regions across the stellar surface.}
        \label{abdor}
\end{figure*}

\begin{figure*}
        \def\subfigtopskip{4pt}
        \def\subfigbottomskip{4pt}
        \def\subfigcapskip{2pt}
        \centering
        \begin{tabular}{cc}
            \subfigure[]{
                        \label{lqhya_periods}
                        \psfig{figure=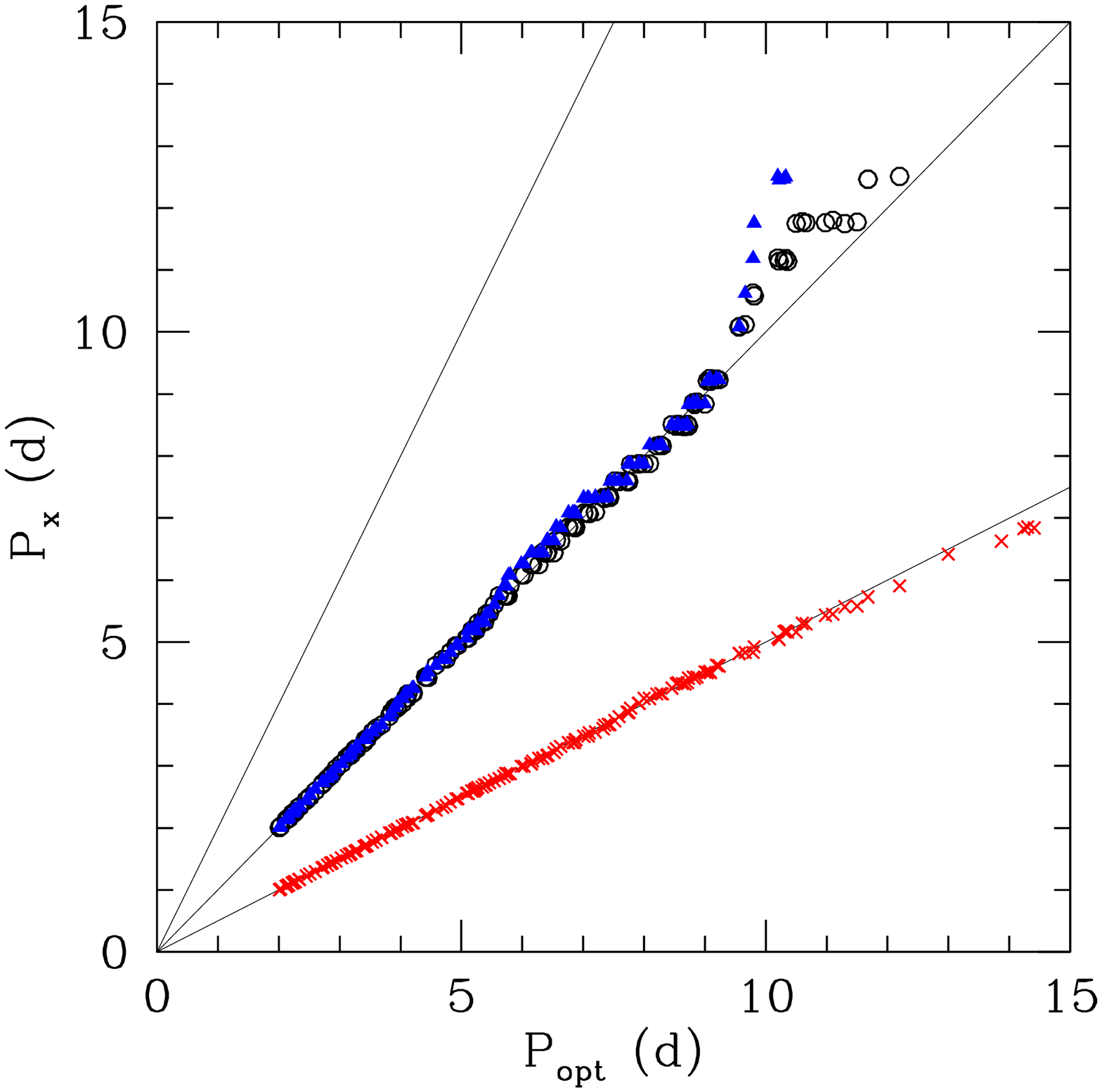,width=80mm}
                        } &
            \subfigure[]{
                        \label{abdor_periods}
                        \psfig{figure=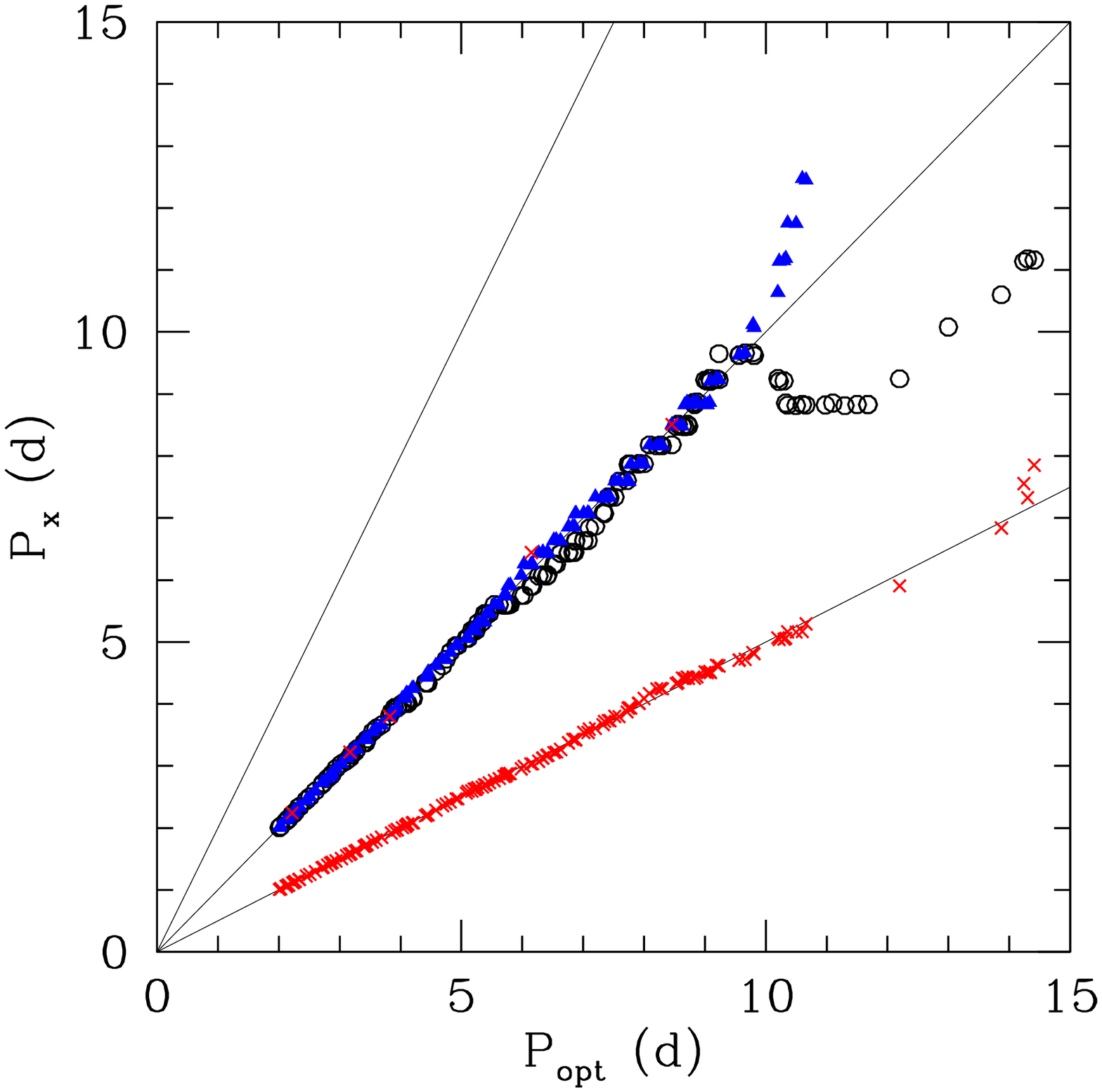,width=80mm}
                        }  \\
        \end{tabular}
        \caption[]{Comparison between X-ray periods $P_X$ and optically determined periods $P_{opt}$ for stars in the 
                   COUP dataset using (a) the LQ Hya-like (b) the AB Dor-like coronal structures shown in Fig. \ref{extrap},
                   for inclinations of $30\degr$({\it black circles}), $60\degr$({\it blue triangles}) and 
                   $90\degr$({\it red crosses}).  The {\it solid lines} represent $P_X=[0.5,1,2]P_{opt}$.  
                   The ratio $P_X/P_{opt}$ is independent of stellar mass and radius.}
        \label{periods}
\end{figure*}

Once we have determined the structure of the closed corona we calculate the 
X-ray emission measure (EM) as described in Appendix A, and plot the ratio of EM to 
the maximum EM against rotational phase.  Fig. \ref{lightcurve} shows such plots 
using both the LQ Hya and AB Dor surface maps, for three inclinations.  The corresponding 
X-ray images are shown in Figs. \ref{lqhya} and \ref{abdor}, from which it can be seen that 
emitting regions are compact and unevenly distributed across the stellar surface.  However, 
some regions (which would contain open field lines) are dark in X-rays, in agreement 
with the conclusions of \citet{fla05}, that the saturation of activity cannot be due
to the filling of the stellar surface with active regions.  

For the LQ Hya-like coronal structure, although there are emitting regions across the stellar surface, 
there are two dominant emitting regions in opposite hemispheres, one of which is brighter in 
X-rays than the other.  The effect of this is most apparent for large inclinations.  For an 
inclination of $i=30\degr$ we see one minimum in the X-ray ``light curve'' 
(see Fig. \ref{lqhya_lightcurve}), however as $i$ is increased to $60\degr$ we see the 
emergence of a second minimum, almost completely out of phase with the first.  For $i=90\degr$ both of
the dominant emitting regions go into eclipse as the star rotates, giving rise to the two distinct 
minima.  As one of these dominant emitting regions is brighter in X-rays, however, the X-ray EM is further
reduced when the brightest region goes into eclipse.  The amplitude of the modulation is around 50\%.  For 
an inclination of $0\degr$ there is no rotational modulation of X-ray emission, as expected.     

The AB Dor-like coronal structure is more complex and, in contrast to the LQ Hya-like coronal structure, 
there are many bright emitting regions (see Fig. \ref{abdor}).  Fig. \ref{abdor_lightcurve} shows
the resulting X-ray light curves, and in this case we see little rotational modulation 
of the X-ray emission for $i=90\degr$, but see modulation of around 50\% for smaller
inclinations.  It is already clear from the shapes of the X-ray light curves that the X-ray period 
depends on the stellar inclination.
 

\subsection{X-ray and optical periods}
\begin{figure*}
        \def\subfigtopskip{4pt}
        \def\subfigbottomskip{4pt}
        \def\subfigcapskip{2pt}
        \centering
        \begin{tabular}{cc}
        \subfigure[]{
                        \label{curve}
                        \psfig{figure=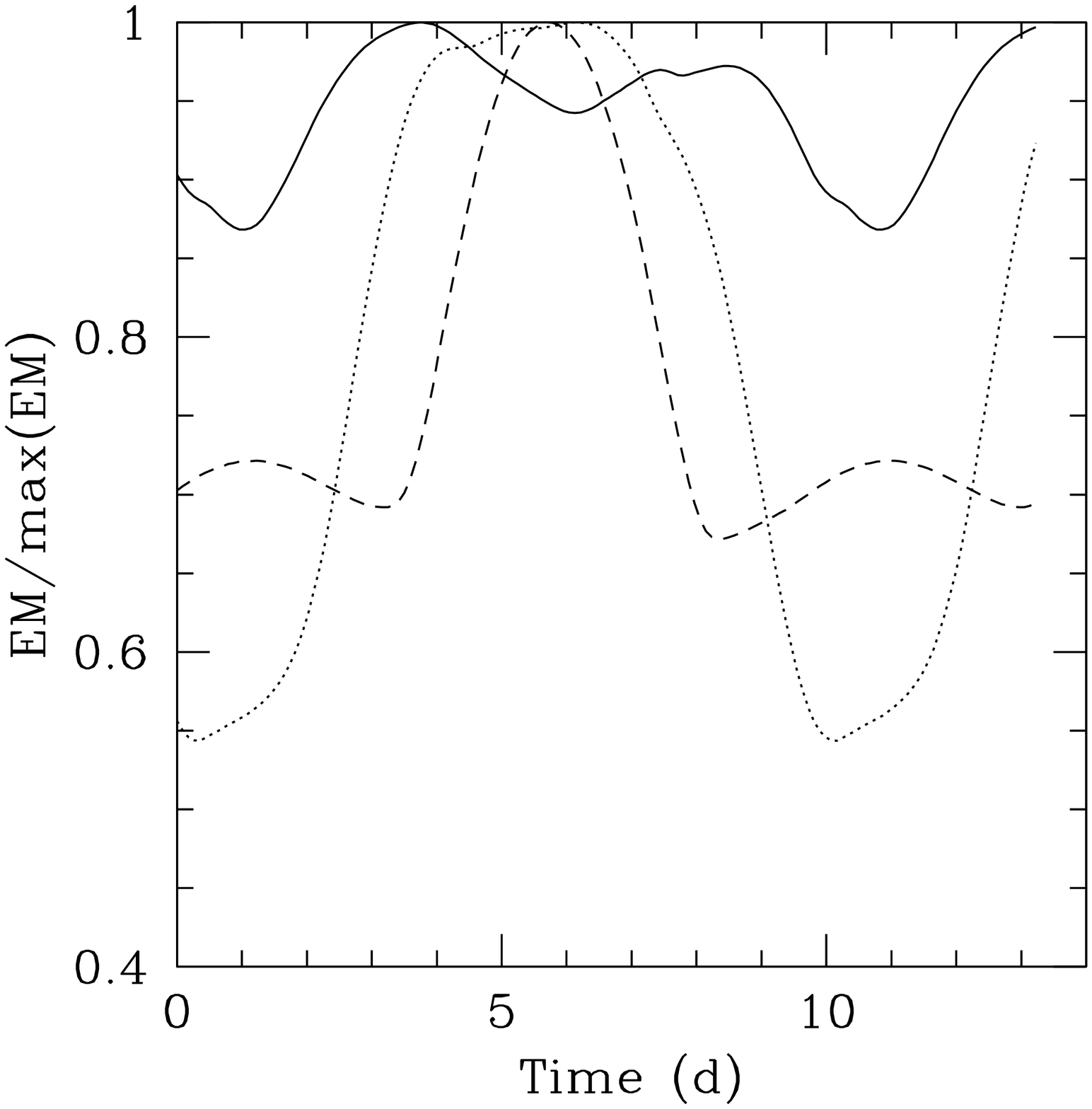,width=80mm}
                        } &
                \subfigure[]{
                        \label{curve_gaps}
                        \psfig{figure=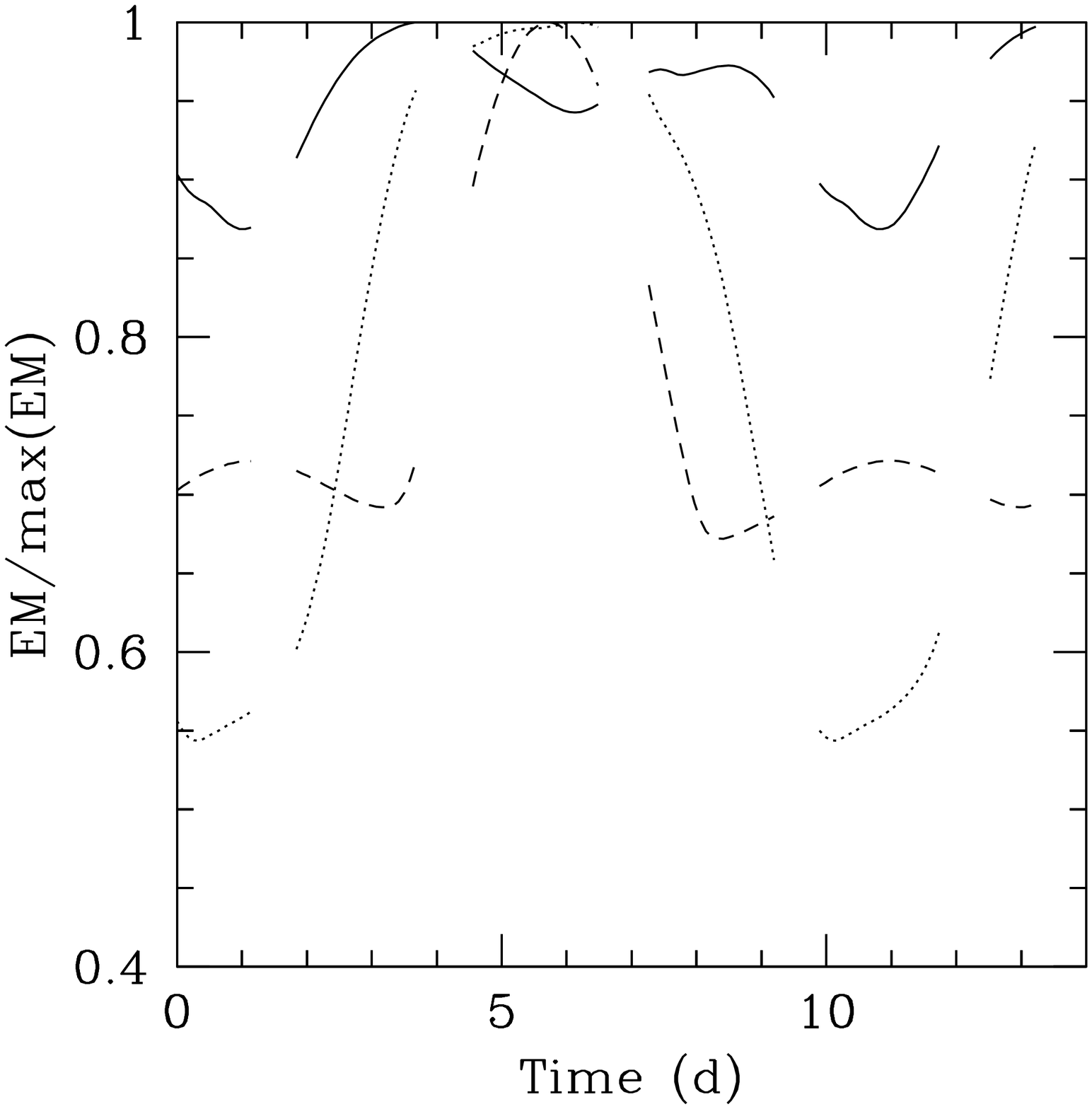,width=80mm}
                        }  \\
                \subfigure[]{
                        \label{lnp}
                        \psfig{figure=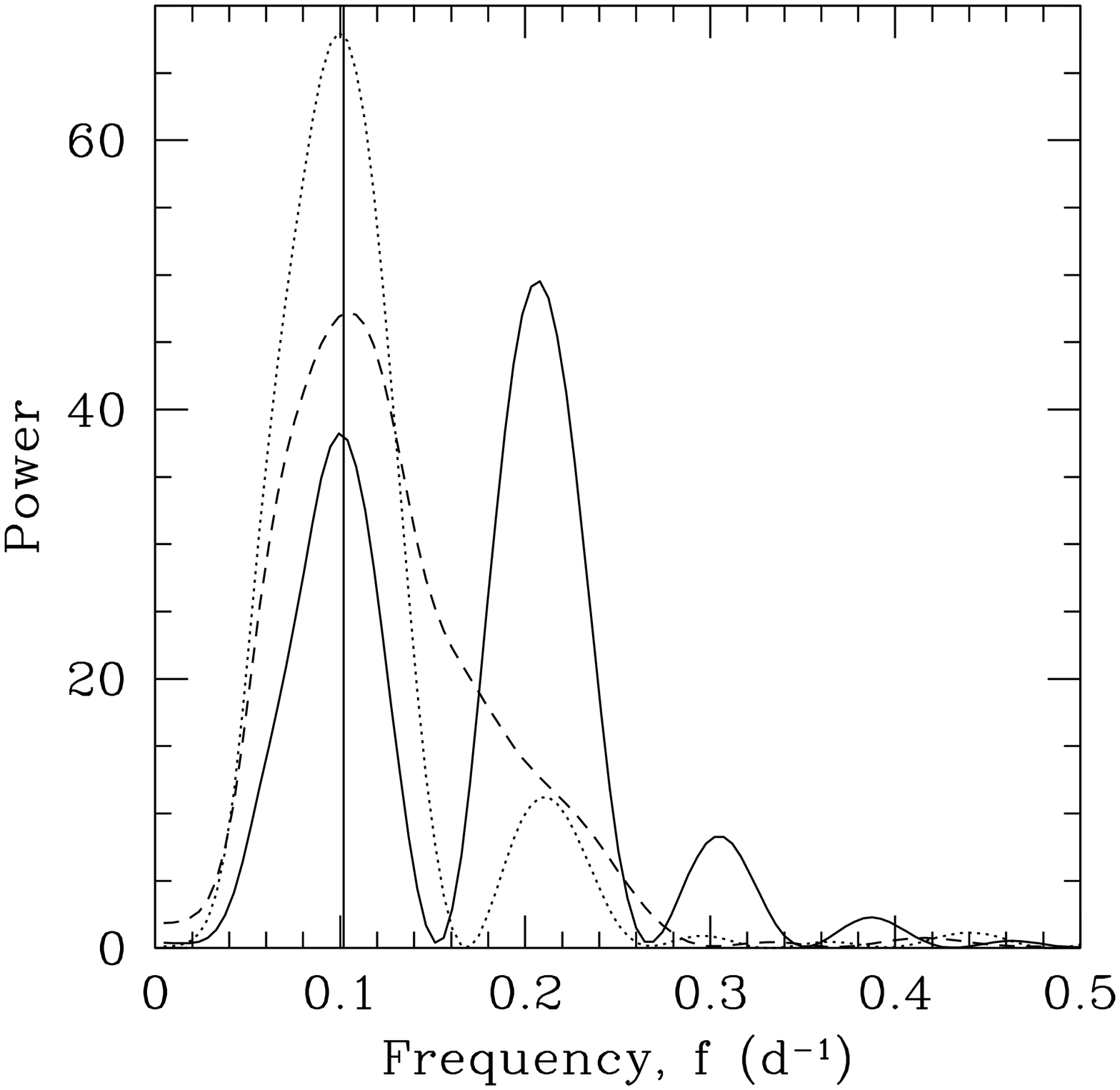,width=80mm}
                        }  &
                \subfigure[]{
                        \label{lnp_gaps}
                        \psfig{figure=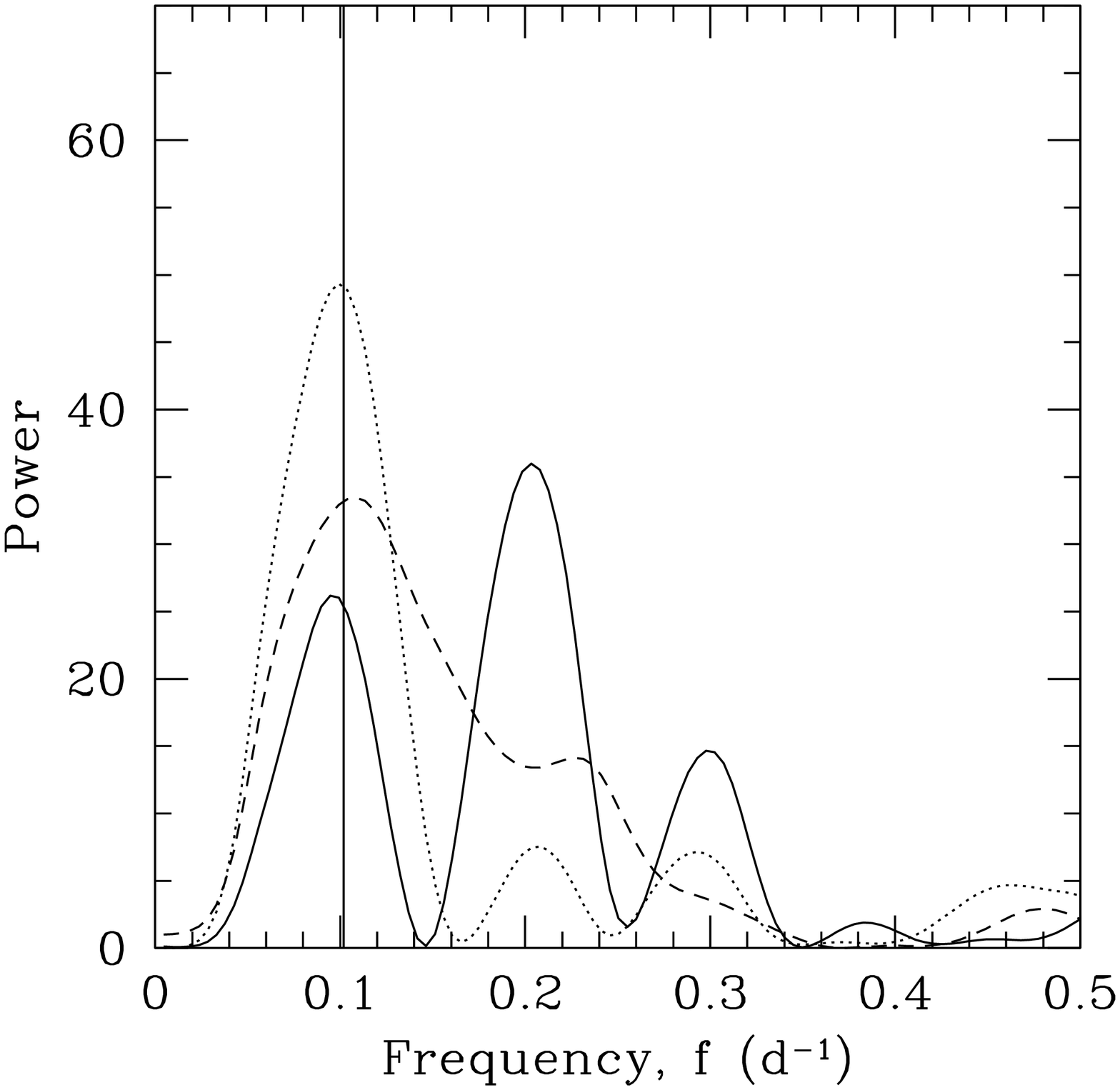,width=80mm}
                        }  \\
        \end{tabular}
        \caption[] {Light curves and Lomb Normalised Periodograms corresponding to inclinations of 30\degr({\it dashed}), 
          60\degr({\it dotted}) and 90\degr({\it solid}), for the coronal structure in Fig. \ref{abdor_extrap}.  The left hand column
          are the simulated light curves without gaps.  However, when gaps are introduced into the light curves (right hand column) we find
          that the most significant peak in the LNP power spectrum generates the same X-ray period, albeit with lower significance.
          There are two significant peaks for the $i=90\degr$ case, one corresponding to $P_{opt}$ and the other $0.5P_{opt}$.  
          The {\it solid vertical line} is the frequency corresponding to the optical period.}
    \label{lnp_curves}
\end{figure*}

We consider the same 233 COUP sources used in the analysis of \citet{fla05}, however, as our model
requires estimates of the stellar parameters (mass, radius, rotation period and measurements from 
which we can infer a coronal temperature) we have omitted those sources without estimates of these
parameters.  This leaves a total of 183 COUP sources with a lower coronal
temperature and 141 where a higher temperature component is also given in the COUP dataset.  For each
COUP source we have generated an X-ray light curve (see Fig. \ref{lnp_curves} for three examples) using
both the LQ Hya and AB Dor surface magnetograms, and for inclinations of $30\degr$, $60\degr$ and
$90\degr$.  In the subsequent discussion we only consider the lower coronal temperature.  Our 
results for the higher coronal temperature are similar and are discussed in \S4.  Each simulated light
curve covers 13.2 days, to match the COUP observation time.  COUP light curves also contain 5 gaps where
the Chandra satellite passed through the Van Allen belts.  Therefore we have considered both
continuous light curves (see Fig. \ref{curve}), and light curves which have gaps at approximately the same
observation times (and lasting for approximately the same duration) as the gaps in COUP X-ray light curves 
(see Fig. \ref{curve_gaps}).  We tried sampling our simulated X-ray light curves every 2000s, 5000s and 10,000s,
to match the analysis of \citet{fla05}, but found that our results were similar in all three cases.  The results
presented here refer only to a time sampling of 10,000s.      

We have calculated X-ray periods using the Lomb Normalised Periodogram (LNP) method so 
that our analysis matches that of \citet{fla05} as closely as possible.
The observed X-ray light curves of COUP sources are contaminated with many flares, with a great
variety of flaring behaviour, whereas our simulated light curves do not suffer from this problem.  
Therefore, when calculating X-ray periods, our false alarms probabilities (FAPs) indicating the significance of 
peaks found in the LNP power spectrum are vanishingly small, with $-40<log_{10}[FAP(\%)]<-10$.    

Fig. \ref{periods} is a plot of our calculated X-ray periods against optically determined
rotation periods from \citet{fla05}, which is collection of rotation periods taken
mainly from \citet{her02}.  We have limited our search to X-ray periods of $P_X < 13d$ and have 
chosen the scale of the axes to allow a direct comparison with similar plots in \citet{fla05}.  
For the LQ Hya-like coronal structure, at any given inclination, the ratio $P_X/P_{opt}$
is independent of stellar mass and radius, except for stars with large $P_{opt}$ (see the discussion below).  
We find that $P_X=P_{opt}$ for inclinations of $30\degr$ and $60\degr$, but that $P_X=0.5P_{opt}$ for $i=90\degr$.  
\citet{fla05} provide a qualitative argument that the X-ray period would equal the optical period if there 
was one dominant emitting region on the stellar surface.  They also point out that if there 
were two dominant emitting regions in opposite hemispheres, separated by $180\degr$ in longitude, 
then it should be expected that the X-ray period would be half of the optical period.  Although 
we agree with this point of view, our LQ Hya-like field structure does have two dominant emitting 
regions in opposite hemispheres, which for high inclinations yields $P_X=0.5P_{opt}$ but at low 
inclinations gives $P_X=P_{opt}$.  Therefore the qualitative picture of having two dominant emitting 
regions in opposite hemispheres giving rise to $P_X=0.5P_{opt}$ should come with the 
(perhaps obvious) caveat that it depends on the inclination at which the star is viewed.  

Results for the AB Dor-like coronal structure are similar (see Fig. \ref{abdor_periods}).  We again find that 
the ratio $P_X/P_{opt}$ is independent of stellar mass and radius for the $i=30\degr$ and $60\degr$ cases, provided that 
$P_{opt} < 10d$.  Stars with $P_{opt}>10d$, and in particular those which have optical periods 
longer than the COUP observing time (13.2d), often do not have X-ray periods of either $0.5P_{opt}$ or $P_{opt}$.
This is most noticeable for the AB Dor-like coronal structure in the $30\degr$ case (see Fig. \ref{abdor_periods}),
where stars with the longest $P_{opt}$ are found to have X-ray periods of $0.7P_{opt}$.  This is a result
of only considering the X-ray light curves over a duration equal to the COUP observing time.  By
extending our simulated X-ray light curves to cover twice the duration of the COUP observing time (so that
$\approx 2$ complete rotations of stars with the longest $P_{opt}$ are covered by the light curves), we found that 
all X-ray periods in the $i=30\degr$ case clustered around $P_X=P_{opt}$.  For the $i=90\degr$ case we find that most stars 
have $P_X=0.5P_{opt}$, but five stars have $P_X=P_{opt}$.  This is due to there being two significant peaks in the LNP power 
spectrum (see Fig. \ref{lnp}) and, depending on which peak is the larger, the X-ray period is either 0.5$P_{opt}$ or
$P_{opt}$.  However, when the AB Dor-like coronal structure is viewed at an inclination of $90\degr$, the amount of rotational 
modulation of X-ray emission is at its smallest (only 13\%) of all the field structures and inclinations which we have 
considered; this can be seen from the light curve in Fig. \ref{abdor_lightcurve}.  Therefore the $i=90\degr$ case for 
the AB Dor-like field is an interesting result.  It shows that it is possible to have a highly structured corona, with emitting regions 
which are close to the stellar surface, with regions which are dark in X-rays, but when such a field structure is viewed at an
unfavourable inclination, little rotational modulation of X-ray emission is detected, and it is difficult to recover an 
X-ray period.  However, there is significant modulation detected for the $i=30\degr$ and $60\degr$ cases with $P_X=P_{opt}$.  
Therefore the X-ray periods we have determined are similar for both the AB Dor and LQ Hya-like coronal fields, and are dependent
on the stellar inclination.  The AB Dor field has many bright emitting regions making it difficult to qualitatively understand 
the exact shape of the resulting X-ray light curve.  In the next section we further investigate the effects of inclination 
and the magnetic field structure on X-ray periods.  

We have also considered how having gaps in our simulated X-ray light curves would affect the resulting X-ray period 
(see Fig. \ref{lnp_curves}).  We found that the values of $P_X$ obtained from light curves with
gaps were identical to those derived from continuous light curves, with the only difference being 
that the main peaks in the LNP power spectra were of lower significance.


\subsection{Effect of inclination and coronal structure}
\begin{figure}
  \centering
  \psfig{file=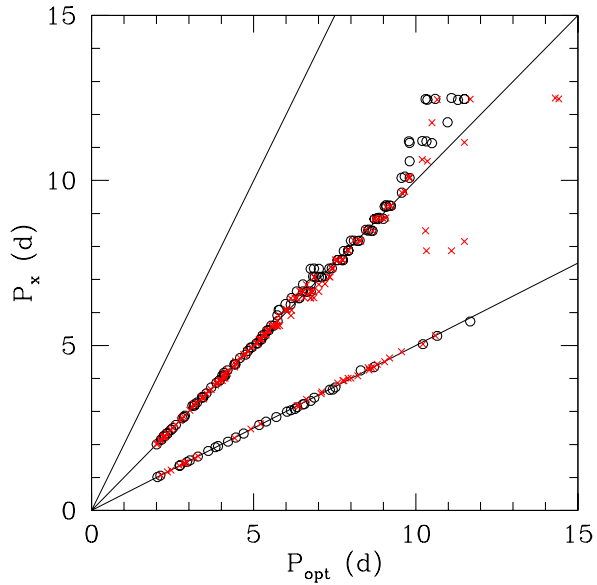,width=80mm}
  \caption{Comparison between our calculated X-ray periods and observed optical periods.  X-ray periods have been calculated for
           COUP stars from \citet{fla05} with randomly assigned inclinations.  The lines represent 
           $P_X=[0.5,1,2]P_{opt}$, with data for the LQ Hya-like ({\it black circles}) and AB Dor-like 
           coronal structures ({\it red crosses}).  There is little difference between the different
           coronal field geometries.}
  \label{random}
\end{figure}

In order to further investigate the effects of inclination and the structure of stellar
coronae, we randomly assign an inclination of between $5\degr$ and $90\degr$ to each of
the COUP sources before calculating an X-ray period.  Inclinations of smaller than 
$\approx 5\degr$ produce little or no rotational modulation of X-ray emission and so 
are not considered.  The results are plotted in Fig. \ref{random} for both the LQ Hya and 
AB Dor-like coronal structures.  We again find that X-ray periods are either equal to $P_{opt}$
or $0.5P_{opt}$, with a few exceptions.    

Our simulations apparently highlight a well defined lower limit for X-ray periods of $P_X=0.5P_{opt}$.  
It should be noted, however, that we are limited by the number of surface maps which we have available; 
it could be the case that another surface magnetogram may yield X-rays periods which are lower
that $0.5P_{opt}$.  \citet{fla05} find stars with reliably detected rotational modulation of X-ray emission with 
$P_X<0.5P_{opt}$ but argue that such period detections may be spurious.  In order
to test if there is a well defined lower limit of $P_X=0.5P_{opt}$ we generated a model coronal
structure which had three emitting regions randomly positioned around the stellar surface.
In the case where all three emitting regions were evenly spaced in longitude, and a fixed (low) latitude
(e.g. the three emitting regions placed at equal increments about the star's equator) we found  
X-ray periods of $P_X=0.333P_{opt}$.  Therefore it is possible to have a coronal field where the 
dominant emitting regions are distributed such that $P_X<0.5P_{opt}$.  However, we find that such field structures 
are rare in comparison to those which yield $P_X=[0.5,1]P_{opt}$.  

\begin{figure*}
        \def\subfigtopskip{4pt}
        \def\subfigbottomskip{4pt}
        \def\subfigcapskip{2pt}
        \centering
        \begin{tabular}{cc}
            \subfigure[]{
                        \label{px_i}
                        \psfig{figure=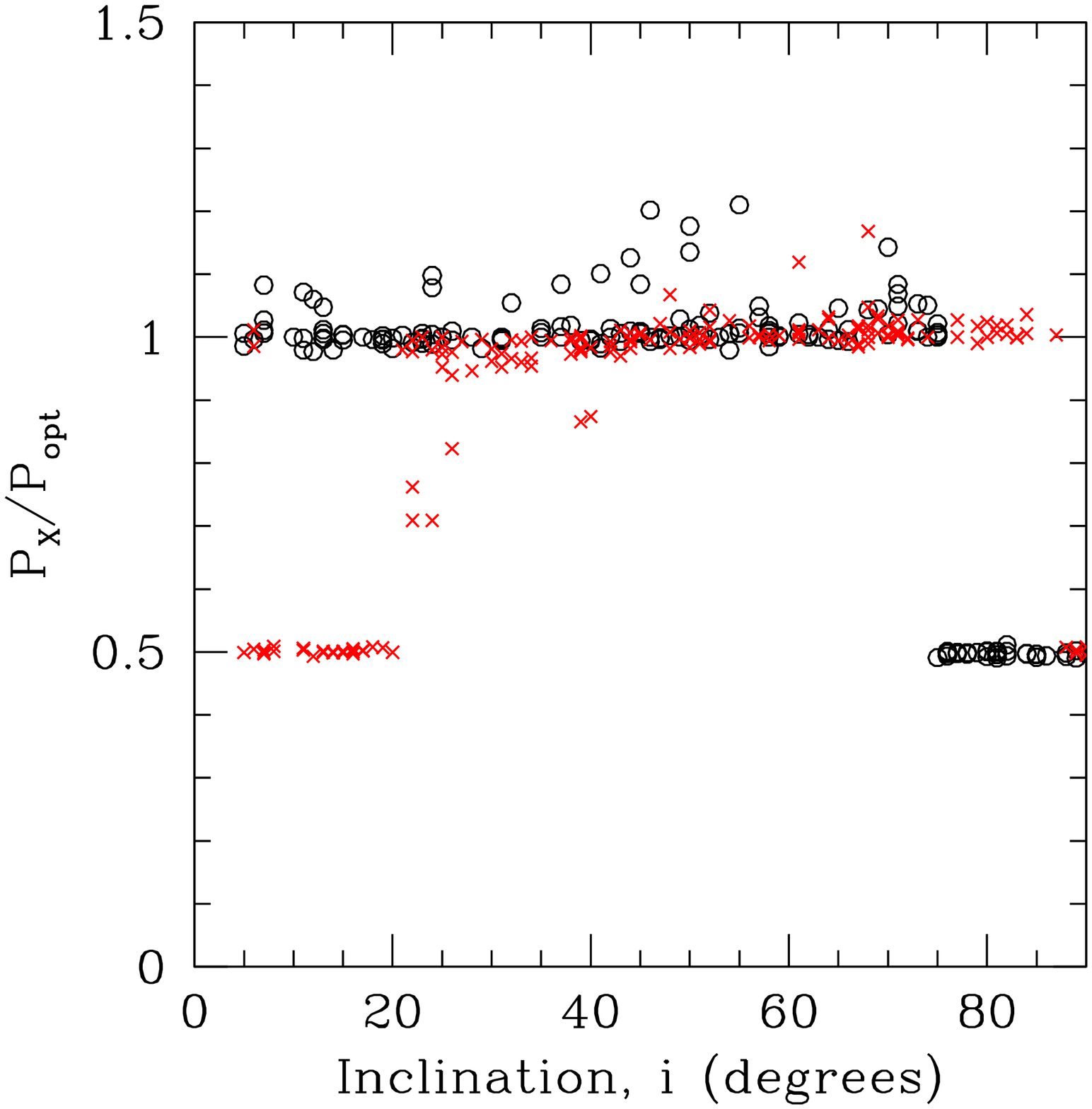,width=80mm}
                        } &
            \subfigure[]{
                        \label{mod_i}
                        \psfig{figure=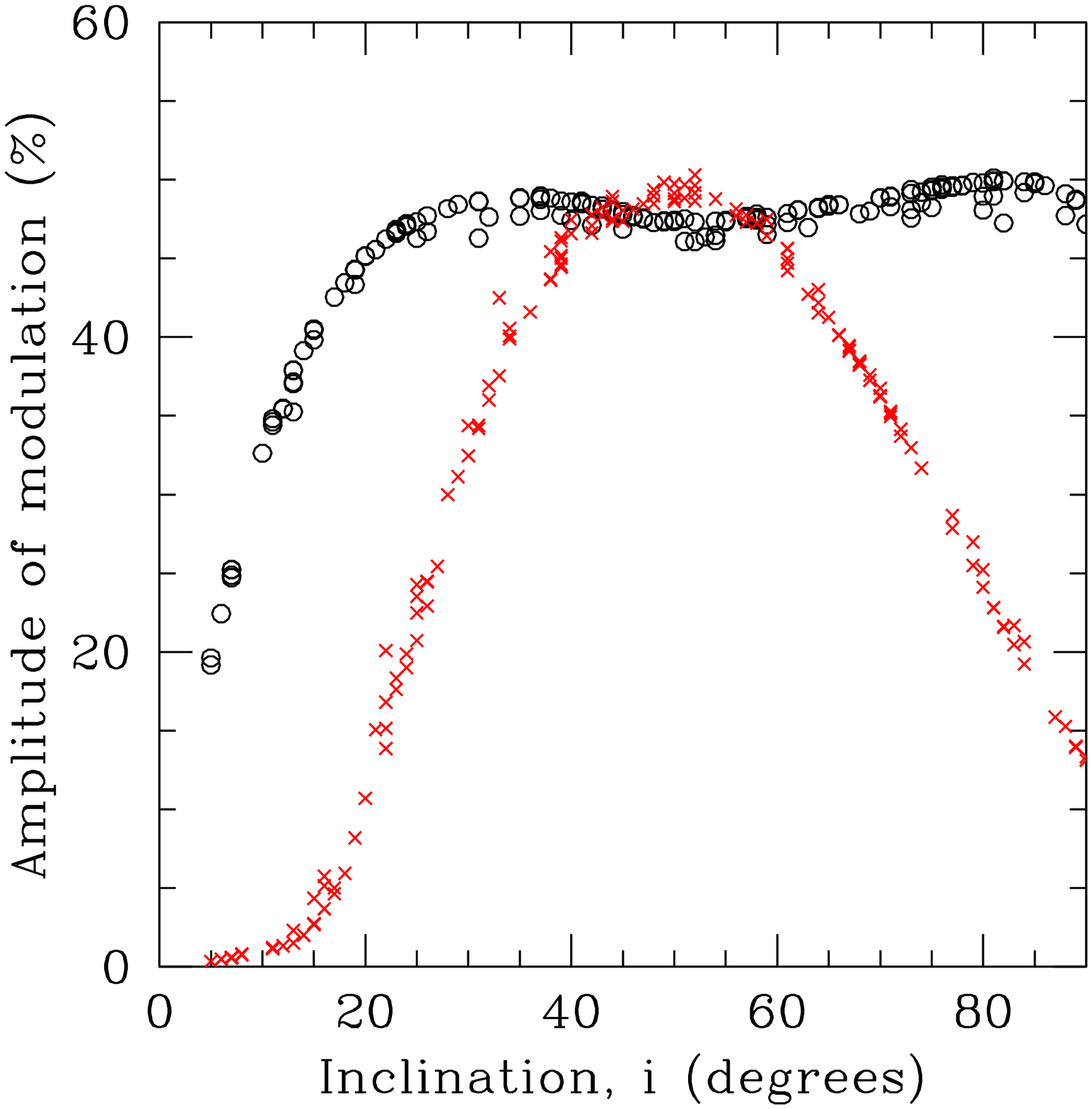,width=80mm}
                        }  \\
        \end{tabular}
        \caption[]{(a) The variation in X-ray period as a fraction of the optical period with stellar inclination for the
                  LQ Hya ({\it black circles}) and AB Dor ({\it red crosses}) coronal fields.  Data represents the same stars
                  as in Fig. \ref{random} with the same randomly selected inclinations, with (b) the amplitude of modulation of 
                  X-ray emission.}
        \label{random_plots}
\end{figure*}

\citet{fla05} also find a few stars with evidence for rotationally modulated X-ray emission with X-ray periods which 
are larger than the optical period, however their further simulations appear to indicate that these period detections
(in particular those with $P_X>2P_{opt}$) are likely to be spurious.  We could not create any model which led to a $P_X$ 
value of much larger than $P_{opt}$.          

The X-ray period depends on stellar inclination (see Fig. \ref{px_i}).  At low inclinations for the 
LQ Hya field, $P_X=P_{opt}$.  As the inclination is increased a second peak in the LNP power spectrum 
becomes stronger and eventually, for $i\approx 75\degr$, both peaks are of almost equal significance.  
For $i>75\degr$, $P_X=0.5P_{opt}$, with the secondary peak having become the most significant.  
This occurs when both of the dominant emitting regions enter eclipse at different times as the star 
rotates.  At low inclinations the amplitude of modulation is small (see Fig. \ref{mod_i}) for the LQ Hya
field, but is almost constant for $i>25\degr$.  Overall we find less modulation for the 
AB Dor field.  This is consistent with our finding from \S3.2.  When the AB Dor field is viewed at both low ($i<30\degr$) 
and high ($i>75\degr$) inclinations there is little rotational modulation of X-ray emission.  Thus, simply having 
compact emitting regions inhomogeneously distributed about the stellar surface is not enough to generate 
large modulation of X-ray emission - the amount of modulation depends strongly on the stellar inclination.
      
\section{Summary and Discussion}
By extrapolating surface magnetograms of young main sequence stars we have shown that
magnetic fields with a realistic degree of complexity reproduce the observed rotational modulation of X-ray emission
from T Tauri stars \citep{fla05}.  This model has already been used to correctly
predict X-ray EMs, mean coronal densities and the observed increase in X-ray EM 
with stellar mass \citep{jar06}.  We find that X-ray emitting regions are 
inhomogeneously distributed around the stellar surface and are typically compact 
($\la R_{\ast}$).  This agrees with the findings of \citet{fla05} who point out that
in order to explain the amplitude of the modulation of X-ray emission which they detect,
emitting regions must be compact in order to undergo eclipse.

We found that the ratio $P_X/P_{opt}$ is independent of stellar mass and radius, 
but that X-ray periods are dependent on stellar inclination.  For example, in the case where there are two 
dominant X-ray emitting regions in opposite hemispheres, we found that the X-ray period could be 
either $0.5P_{opt}$ or $P_{opt}$ depending on how the star is inclined to the observer.  In order words, 
the presence of two bright emitting regions in opposite hemispheres does not immediately imply that 
$P_X=0.5P_{opt}$.  Further, some coronal fields with compact emitting regions produce only 
small amplitude modulation of X-ray emission when viewed at unfavourable inclinations.  

We do not find X-ray periods of $P_X>2P_{opt}$, unlike \citet{fla05}.  However, they argue that those period 
detections are likely to spurious, given the arduous task of removing flares from observed COUP X-ray light curves.  
Also the apparent lower limit of $P_X=0.5P_{opt}$ may not be of true physical origin.  We have demonstrated this by 
constructing a corona which resulted in $P_X=0.333P_{opt}$; however this sort of field structure is perhaps uncommon.  
Instead the majority of X-ray periods cluster around $P_X=[0.5,1]P_{opt}$.  We found that $P_X$ (for a fixed inclination)
was the same for both the LQ Hya and AB Dor-like coronal fields, despite the distribution of emitting regions being different
for the two cases.  This suggests that for complex magnetic fields, where there are 
many emitting regions distributed across the stellar surface,
it becomes difficult to disentangle the contribution to the X-ray emission from any particular emitting region.
Consequently, even though different stars are likely to have different coronal structures, the X-ray periods
of T Tauri stars appear to cluster around $P_{opt}$ and $0.5P_{opt}$.  In other words, although different
coronal structures give rise to different amounts of modulation at a fixed inclination, the X-ray period is 
likely to be the same.         

Throughout we have only considered the lower coronal temperatures derived from the COUP dataset.
If instead we had used the higher coronal temperatures our results and conclusions remain the
same.  For a higher coronal temperature more field lines are unable to contain the coronal
gas and are blown open to form a stellar wind, meaning that coronae are (slightly) more
compact.  This leads to small, but not important, differences in the resulting X-ray periods,
however the amplitude of modulation is larger (typically 60\% compared with 50\% for
the lower coronal temperatures).  The amplitudes of modulation which we determine compare
well with the observed values of \citet{fla05} of 20-70\%.        

Although we have not considered it in this paper, the coronal structure of T Tauri stars may evolve 
on timescales that are comparable with the COUP observing time.  For example a dominant X-ray emitting 
region may be lost due to a large coronal mass ejection which would open up a previously-closed region 
of field.  This may then change the resulting X-ray period.  However, it is still interesting that even
with two different field structures such as those in Fig. \ref{extrap}, the X-ray periods are still typically
equal to the optical period or are half of the optical period.  

We have not considered how the presence of a circumstellar disc and active accretion influences
coronal structures, and the possible resulting effects on the X-ray period/amplitude of modulation.
A comprehensive and detailed discussion of how accretion processes may influence X-ray emission in T Tauri 
stars is provided by \citet{pre05}.  The growing consensus, as was confirmed by COUP, is that X-ray emission
from T Tauri stars does not depend upon the presence of a disc, but is influenced by active accretion.
For example, \citet{fei02} found X-ray activity levels were the same for stars with and without discs\footnote{It is
important to remember that the presence of a disc does not immediately imply that accretion is taking 
place.  Indeed many T Tauri stars show excess infrared emission, indicating the presence of a disc, but lack strong H$\alpha$ 
or CaII emission, characteristic of active accretion.}.  However, \citet{fla03} found a difference in the X-ray
activity levels of accreting and non-accreting stars.  The process of accretion therefore influences the
amount of X-ray emission detected, with stars that show evidence for accretion having X-ray luminosities 
which are a factor of 2-3 smaller than those stars without discs and those stars which show evidence for discs,
but not for accretion \citep{pre05}.  

Accretion may therefore affect the coronal structure of T Tauri stars, which in turn may have implications
for rotational modulation of X-ray emission, and X-ray periods.  It is important to note however that
\citet{fla05} have searched for correlations between observed modulation amplitudes and the $\Delta$(I$-$K) excess, 
a disc indicator, and EW(CaII), an accretion indicator, but do not find any.  Furthermore, \citet{fla05} also 
search for dependencies on the X-ray period detection fraction with the indicators of discs and accretion, but again 
find none.  This suggests that there is no difference in the spatial distribution of X-ray emitting regions 
in both classical and weak line T Tauri stars (with the former having both a disc and active accretion, and the latter
neither). However, \citet{fla05} did find a suggestion (but with a low significance of only about $1\%$) that 
stars with $P_X=0.5P_{opt}$ are preferentially active accretors.  

\citet{jar06} have shown that stars (typically of 
lower mass) which have coronae that would naturally extend to beyond the corotation radius
would have their outer corona stripped by the presence of a disc.  In order to check if this would affect the resulting
X-ray period we determined the coronal structure of all the stars considered in \S3 assuming that they were surrounded
by a disc.  Any field line which passed through the disc at, or within the corotation radius was assumed to have the 
ability to carry an accretion flow and was therefore considered to be ``mass-loaded'' and set to be dark in X-rays - 
a process which is discussed in detail by \citet{jar06} and \citet{gre06}.  We then calculated X-ray emission measures
with rotational phase and determined the amplitude of modulation of X-ray emission and X-ray periods, using the same methods
as discussed in \S3.  We found no difference
in the values of $P_X$ and very little difference in the amplitudes of modulation.  We therefore conclude,
exactly as the observations suggest, that the presence of discs does not influence rotational modulation of X-ray emission.
However, active accretion might.  

We have yet to consider how accretion could influence rotational modulation of X-ray emission.  As already discussed above
stars which are actively accreting show lower levels of X-ray activity than those which are non-accretors.  Again, the possible
explanations for this are discussed by \citet{pre05}, with the most likely being that magnetic reconnection
events in the magnetosphere cannot heat the dense material within accretion columns to a high enough temperature to emit
in X-rays.  Some models suggest that accretion will distort magnetic field lines giving rise to instabilities and reconnection
events (e.g. \citealt{rom04}).  The energy liberated in such reconnection events may not be able to heat the high density
material in accretion columns to a sufficiently high temperature in order to emit in X-rays.  Thus accretion columns rotating 
across the line-of-sight to 
the star may be responsible for the observed reduction in X-ray emission in accreting T Tauri stars compared
with the non-accretors.  The effect of this on X-ray periods and modulation amplitudes will be considered in future work.
A detailed accretion flow model will have to be constructed in order to account for the attenuation of coronal X-rays by
accretion columns rotating across the line-of-sight.  Also, inner disc warps may have a similar effect at high stellar inclinations.
This could change the shape of the X-ray variability curves shown in Figs. \ref{lightcurve} and \ref{lnp_curves}, with a 
reduction in X-ray emission at particular rotational phases, depending on the accretion flow geometry.

Our model may allow for the simultaneous confinement of X-ray emitting plasma in closed loops 
close to the stellar surface, but also for other field lines to stretch out into the inner disc and carry
accretion flows.  \citet{gre06} find that often open field lines can carry accretion flows, which
may allow accretion to influence the magnetic field structure on the large scale, with field lines being 
``mass-loaded'' with disc material,  but leaving compact coronae relatively unaffected by the process of 
accretion.  This model would allow for the existence of large extended magnetic structures, inferred to exist from
the detection of large flaring loops by \citet{fav05}, with the coexistence of compact coronae.  However, much 
quantitative work remains to be done here.  Another new accretion 
model may offer insights into the connection between X-ray emission and accretion in T Tauri stars.   
An MHD model by \citet{von06} suggests that rather than material being loaded onto field lines
at the inner edge of the disc, it flows onto the star only at times when, and into regions where, the accretion
flow is able to overcome the stellar wind.  In such a way this model predicts that accretion only occurs into
regions of low field strength.  It is also worth noting the observational results of 
\citet{fla05} that indicate that rotational modulation of X-ray emission is not preferentially detected in either 
accreting or non-accreting T Tauri stars and that therefore the spatial structure of the X-ray emitting plasma is the same in both 
cases.  Hence, although accretion does affect X-ray emission in accreting stars, and may distort the structure
of stellar coronae, it may not affect the existence of rotational modulation.


\section*{Acknowledgements}
The authors thank Add van Ballegooijen who wrote the original version of the 
potential field extrapolation code and Keith Horne for valuable discussions.  SGG 
acknowledges the support from a PPARC studentship.
\bibliographystyle{mn2e}
\bibliography{xrays}

\begin{thebibliography}{}

\bibitem[\protect\citeauthoryear{{Altschuler} \& {Newkirk}}{{Altschuler} \&
  {Newkirk}}{1969}]{alt69}
{Altschuler} M.~D.,  {Newkirk} G.,  1969, SoPh, 9, 131

\bibitem[\protect\citeauthoryear{{Donati}}{{Donati}}{1999}]{don99a}
{Donati} J.-F.,  1999, MNRAS, 302, 457

\bibitem[\protect\citeauthoryear{{Donati} \& {Collier Cameron}}{{Donati} \&
  {Collier Cameron}}{1997}]{don97a}
{Donati} J.-F.,  {Collier Cameron} A.,  1997, MNRAS, 291, 1

\bibitem[\protect\citeauthoryear{{Donati}, {Collier Cameron}, {Hussain} \&
  {Semel}}{{Donati} et~al.}{1999}]{don99b}
{Donati} J.-F.,  {Collier Cameron} A.,  {Hussain} G.~A.~J.,    {Semel} M.,
  1999, MNRAS, 302, 437

\bibitem[\protect\citeauthoryear{{Donati} {et al.}}{{Donati} {et
  al.}}{2003}]{don03}
{Donati} J.-F. {et al.} 2003, MNRAS, 345, 1145

\bibitem[\protect\citeauthoryear{{Donati}, {Semel}, {Carter}, {Rees} \&
  {Collier Cameron}}{{Donati} et~al.}{1997}]{don97b}
{Donati} J.-F.,  {Semel} M.,  {Carter} B.~D.,  {Rees} D.~E.,    {Collier
  Cameron} A.,  1997, MNRAS, 291, 658

\bibitem[\protect\citeauthoryear{{Favata}, {Flaccomio}, {Reale}, {Micela},
  {Sciortino}, {Shang}, {Stassun} \& {Feigelson}}{{Favata}
  et~al.}{2005}]{fav05}
{Favata} F.,  {Flaccomio} E.,  {Reale} F.,  {Micela} G.,  {Sciortino} S.,
  {Shang} H.,  {Stassun} K.~G.,    {Feigelson} E.~D.,  2005, ApJS, 160, 469

\bibitem[\protect\citeauthoryear{{Feigelson}, {Broos}, {Gaffney} III,
  {Garmire}, {Hillenbrand}, {Pravdo}, {Townsley} \& {Tsuboi}}{{Feigelson}
  et~al.}{2002}]{fei02}
{Feigelson} E.~D.,  {Broos} P.,  {Gaffney} III J.~A.,  {Garmire} G.,
  {Hillenbrand} L.~A.,  {Pravdo} S.~H.,  {Townsley} L.,    {Tsuboi} Y.,  2002,
  ApJ, 574, 258

\bibitem[\protect\citeauthoryear{{Flaccomio}, {Damiani}, {Micela}, {Sciortino},
  {Harnden} Jr., {Murray} \& {Wolk}}{{Flaccomio} et~al.}{2003}]{fla03}
{Flaccomio} E.,  {Damiani} F.,  {Micela} G.,  {Sciortino} S.,  {Harnden} Jr.
  F.~R.,  {Murray} S.~S.,    {Wolk} S.~J.,  2003, ApJ, 582, 398

\bibitem[\protect\citeauthoryear{{Flaccomio}, {Micela}, {Sciortino},
  {Feigelson}, {Herbst}, {Favata}, {Harnden} \& {Vrtilek}}{{Flaccomio}
  et~al.}{2005}]{fla05}
{Flaccomio} E.,  {Micela} G.,  {Sciortino} S.,  {Feigelson} E.~D.,  {Herbst}
  W.,  {Favata} F.,  {Harnden} F.~R.,    {Vrtilek} S.~D.,  2005, ApJS, 160, 450

\bibitem[\protect\citeauthoryear{{Getman} {et al.}}{{Getman} {et
  al.}}{2005}]{get05}
{Getman} K.~V. {et al.} 2005, ApJS, 160, 319

\bibitem[\protect\citeauthoryear{{Giardino}, {Favata}, {Silva}, {Micela},
  {Reale} \& {Sciortino}}{{Giardino} et~al.}{2006}]{gia06}
{Giardino} G.,  {Favata} F.,  {Silva} B.,  {Micela} G.,  {Reale} F.,
  {Sciortino} S.,  2006, A\&A, 453, 241

\bibitem[\protect\citeauthoryear{{Gregory}, {Jardine}, {Simpson} \&
  {Donati}}{{Gregory} et~al.}{2006}]{gre06}
{Gregory} S.~G.,  {Jardine} M.,  {Simpson} I.,    {Donati} J.-F.,  2006, MNRAS,
  371, 999

\bibitem[\protect\citeauthoryear{{Herbst}, {Bailer-Jones}, {Mundt},
  {Meisenheimer} \& {Wackermann}}{{Herbst} et~al.}{2002}]{her02}
{Herbst} W.,  {Bailer-Jones} C.~A.~L.,  {Mundt} R.,  {Meisenheimer} K.,
  {Wackermann} R.,  2002, A\&A, 396, 513

\bibitem[\protect\citeauthoryear{{Jardine}, {Cameron}, {Donati}, {Gregory} \&
  {Wood}}{{Jardine} et~al.}{2006}]{jar06}
{Jardine} M.,  {Cameron} A.~C.,  {Donati} J.-F.,  {Gregory} S.~G.,    {Wood}
  K.,  2006, MNRAS, 367, 917

\bibitem[\protect\citeauthoryear{{Jardine}, {Collier Cameron} \&
  {Donati}}{{Jardine} et~al.}{2002a}]{jar02a}
{Jardine} M.,  {Collier Cameron} A.,    {Donati} J.-F.,  2002a, MNRAS, 333, 339

\bibitem[\protect\citeauthoryear{{Jardine}, {Wood}, {Collier Cameron}, {Donati}
  \& {Mackay}}{{Jardine} et~al.}{2002b}]{jar02b}
{Jardine} M.,  {Wood} K.,  {Collier Cameron} A.,  {Donati} J.-F.,    {Mackay}
  D.~H.,  2002b, MNRAS, 336, 1364

\bibitem[\protect\citeauthoryear{{Preibisch} {et al.}}{{Preibisch} {et
  al.}}{2005}]{pre05}
{Preibisch} T. {et al.} 2005, ApJS, 160, 401

\bibitem[\protect\citeauthoryear{{Romanova}, {Ustyugova}, {Koldoba} \&
  {Lovelace}}{{Romanova} et~al.}{2004}]{rom04}
{Romanova} M.~M.,  {Ustyugova} G.~V.,  {Koldoba} A.~V.,    {Lovelace} R.~V.~E.,
   2004, ApJL, 616, L151

\bibitem[\protect\citeauthoryear{{Valenti} \& {Johns-Krull}}{{Valenti} \&
  {Johns-Krull}}{2004}]{val04}
{Valenti} J.~A.,  {Johns-Krull} C.~M.,  2004, Ap\&SS, 292, 619

\bibitem[\protect\citeauthoryear{{van Ballegooijen}, {Cartledge} \&
  {Priest}}{{van Ballegooijen} et~al.}{1998}]{van98}
{van Ballegooijen} A.~A.,  {Cartledge} N.~P.,    {Priest} E.~R.,  1998, ApJ,
  501, 866

\bibitem[\protect\citeauthoryear{{von Rekowski} \& {Piskunov}}{{von Rekowski}
  \& {Piskunov}}{2006}]{von06}
{von Rekowski} B.,  {Piskunov} N.,  2006, Astron. Nachr., 327, 340

\end{thebibliography}


\appendix

\section[]{Determining X-ray Emission Measures} 
In deriving expressions for the variation of the emission measure with 
radial distance from the star, we begin with (\ref{phydro}). If we scale all distances 
to a stellar radius, we can write
\begin{equation}
 p(r,\theta) = p_\ast(1,\theta)\exp\left[ \Phi_g\left(\frac{1}{r} -1\right) + \Phi_c\sin^2\theta(r^2-1)\right]
\end{equation}
where $p_\ast(1,\theta)$ is the pressure distribution across the stellar surface and 
\begin{eqnarray}
 \Phi_g & = & \frac{GM/R_\ast}{k_BT/m} \\
 \Phi_c & = & \frac{\omega^2R_\ast^2/2}{k_BT/m}
\end{eqnarray}
are the surface ratios of gravitational and centrifugal energies to the thermal energy. 
The emission measure $EM = \int n_e^2 dV$ is then
\begin{eqnarray}
 EM & = & \frac{4\pi R_\ast^3}{(k_B T)^2} \\ \nonumber
        &    & \int_{\theta_m}^{\pi/2}
       p_\ast^2(1,\theta)\exp \left [-2\Phi_c \cos^2 \theta (r^2-1)\right] \sin\theta d\theta \\ \nonumber
        &    &  \int_{1}^{r_m}r^2\exp \left[2\Phi_g \left(\frac{1}{r}-1\right) + 2 \Phi_c(r^2 -1) \right] dr. \nonumber
\end{eqnarray}
With the substitution $\mu = \cos \theta$ such that 
$\mu_m = \cos \theta_m = (1-r/r_m)^{1/2}$, the $\theta$-integral can be written as 
\begin{equation}
 \int_{0}^{\mu_m}p_\ast^2(1,\theta)\exp \left[ -2\Phi_c \mu^2 (r^2-1) \right] d\mu
\end{equation}
which, on using the substitution $t=[2\Phi_c (r^2-1)]^{1/2} \mu $, gives
\begin{equation}
 \frac{1}{\left[2\Phi_c (r^2-1)\right]^{1/2}} \int_0^{t_m} p_\ast^2(1,\theta) e^{-t^2} dt
\end{equation}
where $t_m = [2\Phi_c (r^2-1)]^{1/2} [1-r/r_m]^{1/2}$.  Assuming that the pressure is 
uniform over the stellar surface we obtain the following expression for the emission measure 
as a function of $r$:
\begin{eqnarray}
  EM(r) & = & \sqrt{ \frac{2\pi^3} {\Phi_c} }\frac{R_\ast^3 p_\ast^2}{(k_B T)^2} \\ \nonumber
        &    &\int_{1}^{r_m}
               \frac{r^2}{(r^2 - 1)^{1/2}} {\rm erf} \left[2\Phi_c(r^2-1)\left(1-\frac{r}{r_m}\right) \right]^{1/2}\\ \nonumber
         &  &      \exp\left[2\Phi_g  \left(\frac{1}{r}-1\right) + 2 \Phi_c(r^2 -1)\right] dr. \\ \nonumber
\end{eqnarray}
where we remind the reader that all distances $r$ have been scaled to a stellar radius.

\bsp

\label{lastpage}

\end{document}